IAC-21-66537

# A Portuguese radar tracking sensor for Space Debris monitoring

João Pandeirada[a,b]*, Miguel Bergano[a], Paulo Marques[c], Domingos Barbosa[a], Bruno Coelho[a], Valério Ribeiro[a], José Freitas[d], Domingos Nunes[a], José Eduardo[a,b]

[a] *Instituto de Telecomunicações, Aveiro, Portugal;* joao.pandeirada@av.it.pt (J.P); jbergano@av.it.pt (M.B); dbarbosa@av.it.pt (D.B); brunodfcoelho@av.it.pt (B.C); valerio.ribeiro@av.it.pt (V.R); dfsn@av.it.pt (D.N); jecs@av.it.pt (J.E);
[b] *Department of Electronics, Telecommunications and Informatics, University of Aveiro, Aveiro, Portugal*
[c] *Instituto de Telecomunicações / ISEL-IPL, Lisboa, Portugal;* pmarques@isel.pt (P.M)
[d] *Ministério da Defesa Nacional, Portugal;* jose.freitas@defesa.pt (J.F)
\* Corresponding Author

**Abstract**

The increase in space debris is a threat to space assets, space based-operations and led to a common effort to develop programs for dealing with this increase. As part of the Portuguese Space Surveillance and Tracking (SST) project, led by the Portuguese Ministry of Defense (MoD), the Instituto de Telecomunicações (IT) is developing rAdio TeLescope pAmpilhosa Serra (ATLAS), a new monostatic radar tracking sensor located at the Pampilhosa da Serra Space Observatory (ErPoB), Portugal. The system operates at 5.56 GHz and aims to provide information on objects in low earth orbit (LEO) orbits, with cross sections above 10 cm$^2$ at 1000 km. ErPoB houses all the necessary equipment to connect to the research and development team in IT-Aveiro and to the European Union Space Surveillance and Tracking (EU-SST) network through the Portuguese SST-PT network and operation center. The ATLAS system features digital waveform synthesis, power amplifiers using Gallium Nitride (GaN) technology, fully digital signal processing and a highly modular architecture that follows an Open Systems (OS) philosophy and uses Commercial-Off-The-Shelf (COTS) technologies. ATLAS establishes a modern and versatile platform for fast and easy development, research and innovation. The whole system (except antenna and power amplifiers) was tested in a setup with a major reflector of opportunity at a well defined range. The obtained range profiles show that the target can be easily detected. This marks a major step on the functional testing of the system and on getting closer to an operational system capable of detecting objects in orbit.

**Acronyms/Abbreviations**

National Operation Center (NOC), Tracking Data Message (TDM), Low Earth Orbit (LEO), Low Noise Amplifier (LNA), Signal to Noise Ratio (SNR), Radar Cross Section (RCS), Constant False Alarm Rate (CFAR)

## 1. Introduction

Since the beginning of the space age in the 1950's, humanity has been deploying various systems to the orbital environment. This trend will certainly continue to grow due to the increasing number of space objects [1] and the entry of the "new space" stakeholders and mega constellations.

Space-based services, such as communications, earth observation, navigation and timing provide essential support for economic and social well-being, for public safety and for the functioning of key government responsibilities [2]. These are now at an increasing risk of collision in space with other operational spacecraft or debris. In order to mitigate these risks, we require the capacity to survey and track all the objects to provide timely and accurate information to the different stakeholders. SST networks provide a constant monitoring of space-assets in order to predict their orbits and ultimately avoid collisions. These networks are mainly constituted by ground-based optical sensors and radar systems which are responsible for measuring the state vector of orbiting objects.

The EU-SST is a support framework for the creation of an autonomous monitoring network at the European level that contributes to the detection and tracking of space debris and issues alerts when actions are necessary [3]. Portugal is a member of the EU-SST program and is developing capabilities both in optical and radar sensors.

This article showcases ATLAS, a new radar for space debris detection and tracking that is being developed in Portugal [4] as well as the whole infrastructure surrounding it. In Section 2 we introduce the radar system with a brief technical description of all the components and explain how the interfacing between ATLAS, IT and the EU-SST is established. In Section 3 we take the opportunity to show the most recent functional tests performed with the system before it is deployed for operation. In Section 4 we discuss the results obtained from the experimental setup presented in Section 3. Finally, in Section 5 we present the current status of the work and the next steps in order to have a fully functional system.

## 2. ATLAS System and Infrastructure
### 2.1 Infrastructure

IT is a Linked Third-Party to the MoD in the EU-SST program and initiated in 2019 a major upgrade of its 9-meter Cassegrain Antenna located at the Pampilhosa da Serra Space Observatory (ErPoB), Portugal. This upgrade is a ground-based radar system, named ATLAS, for tracking objects above 10 $cm^2$ at 1000 km radar range. The main SST service that ATLAS will provide to the EU-SST network is to update orbital information for all detected objects, which will ultimately lead to a more precise and up to date object catalogue.

The ErPoB station currently houses a smaller 5 m antenna used for radio astronomy studies, an optical sensor operated by the MoD for SST, a weather station, the ATLAS antenna, motors, motor controllers, security cameras and a server. The radar system itself will be deployed as soon as all the components are fully tested and delivered. All these subsystems are connected to an operations center which is connected to a dedicated high-speed fiber connection. This allows remote access to the whole system through a secure connection to IT-Aveiro. The system is supported by the EngageSKA [5] infrastructure which has a dedicated team to maintain and operate the whole network, operate ATLAS as a service for SST-PT and use the system as a platform for research and development on advanced signal processing techniques for radar.

Fig. 1 illustrates the complete infrastructure regar--ding the ATLAS system. The ErPoB station contains an operation center with all the software and hardware necessary to interface with the radar system and with the outside world. The operations center also feeds an internal object catalogue, used for tests and calibration purposes.

The Portuguese NOC, which interfaces with the EU-SST network, is located in the Azores island. The Azores NOC will be responsible for tasking ATLAS with tracking requests and ATLAS should return TDM files [6]. TDM files consist of up to date information about the orbit of the objects and are used by the NOC to improve the national catalogue. The object catalogues from all nations that participate in the EU-SST are then used to update the European catalogue.

### 2.2 Radar System

ATLAS is a monostatic radar using solid state power amplifiers (GaN) with a peak output power of 5 kW for tracking debris objects in LEO. Regarding the receiver side, the system is coherent with detection and processing fully in digital domain with a bandwidth of 50 MHz and with capacity of detecting Doppler velocities up to 10.79 km/s. The system follows the use of COTS technologies and OS which enables a major cost reduction in the system development, maintenance and future upgrade [7]. ATLAS characteristics are summarized in Table 1 and some of its components are depicted in Fig. 2. As a tracking sensor, it uses a narrow beamwidth antenna which enables precise measurements of single objects' position and velocity as opposed to surveillance radars which are designed to cover larger portions of the sky.

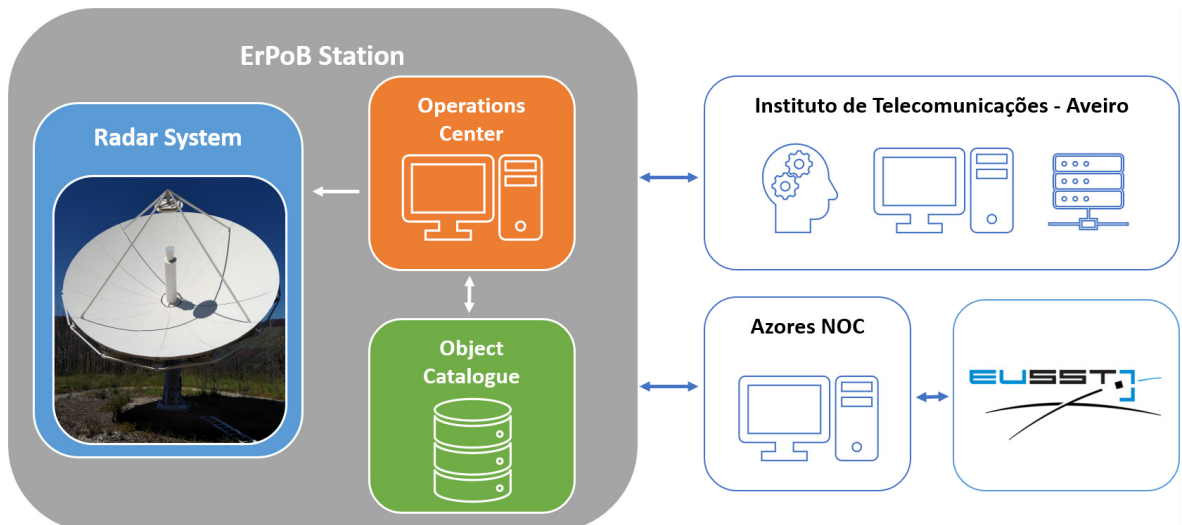

Fig. 1. ATLAS Infrastructure.

In order to have a coherent receiver, the coherence between the transmitted and received signals needs to be maintained after several milliseconds (which corresponds to distances of several hundred kilometers). This requires an enormous phase noise (and phase/frequency) stability of the signal generated, which led to the design of a highly precise frequency synthesizer with a low phase noise [8].

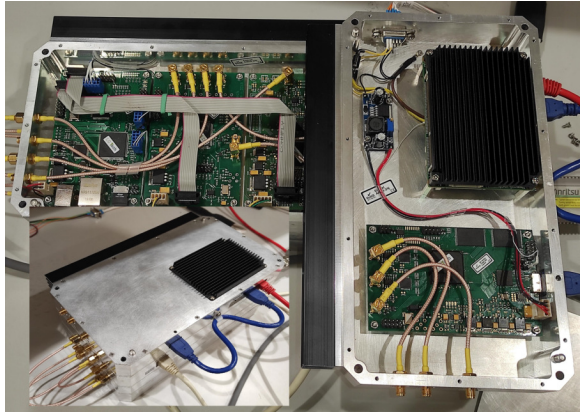

Fig. 2. ATLAS back-end components: acquisition board (right), controller board (top-left), complete system (bottom-left).

Coherent detection leads to efficient pulse integration which, together with the low noise figure of the LNA and high antenna gain results in the SNR shown in Table 1.

| Table 1. ATLAS Features | |
|---|---|
| Transmitter Frequency | 5.56 GHz |
| Antenna Gain | 46 dB |
| Antenna Beamwidth | 44 arcmin |
| Peak power Transmitter PA | 5 kW |
| Technology | Solid State GaN |
| Waveform | Arbitrary amplitude modulation |
| Max. Pulse length | 10 s |
| Pulse repetition frequency | 10 MHz (max) |
| Phase noise | -91.3 dBc[Hz] @ 100 kHz |
| Intermediate frequency | 400 MHz |
| LNA Noise Figure | 0.7 dB |
| Receiver type | Coherent Receiver |
| Receiver Bandwidth | 50 MHz |
| Back-end processing | Base-band I & Q complex data acquisition (local DSP, OS Linux) |
| Data pipeline | Processed level-1 files created automatically to be retrieved by remote connection SFTP. |
| SNR for 1 $m^2$ RCS at $10^3$ km | 39.55 dB |
| SNR for 10 $cm^2$ RCS at $10^3$ km | 9.55 dB |

With arbitrary waveform design in the digital domain, the emitted signal can be tailored to the predominant scatterers of the targets, resulting in a better matched filtering which ultimately leads to better SNR and pulse compression [9]. Also, by having 50 MHz of bandwidth both in the transmitter and receiver, ATLAS allows the usage and testing of waveforms with extremely high compression factors, making it possible to obtain distance measurements with error below 0.05% at 1000 km [4]. ATLAS allows the design of a dynamic signal processing chain that optimizes the quality of the received data by measuring and modelling the processes that affect the characteristics of the received echoes [10].

Due to the well known constraints imposed by the COVID-19 pandemic, especially in the semiconductor industry, the power amplifier modules are still in development. The rest of the system is in final tweaking and testing.

### 3. End-to-end tests
*3.1 Experimental Setup*

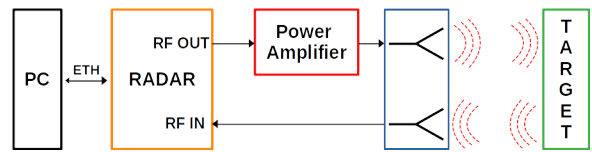

Fig. 3. Experimental setup diagram.

ATLAS is currently under functional testing in order to be fully characterized before being installed on the antenna at the ErPoB. One of the most important functional tests in radar systems are the end-to-end tests which consist of preparing an outdoor setup with well defined targets and then using the radar to measure the distance to those targets. This allows testing of the whole system from the waveform synthesis to the reception and subsequent signal processing. By doing this in a well controlled environment, it is possible to minimize and avoid mistakes as well as predict future operational problems that are much harder to debug when pointing to orbiting objects.

Figure 3 shows the experimental setup used for the end-to-end testing. The radar emitter/receiver were connected each to two identical horn antennas that were aligned and spaced 1 meter apart. These antennas are not the ones described in Table 1, they are horn antennas developed specifically for this purpose and have a gain of 17 dB and a beamwidth of 12º. The emitter antenna was connected to the radar with a 5 W power amplifier. The radar was controlled and configured by a PC through an Ethernet connection. The PC configures the complete radar from the pulse generation to the acquisition. The PC triggers the radar to emit a burst of pulses which will be echoed by a target and captured by the receiver antenna. The acquisition system will generate several files that are processed by the PC to calculate the distance to the target. A photo of the setup is shown in Fig. 4 where it is possible to identify the antennas in the blue box and the power amplifier and radar in the red and orange boxes, respectively. The green box corresponds to a big metallic structure on the highway and was used as a target. The purple box corresponds to buildings that are candidates for other reflections that will be analyzed in the results.

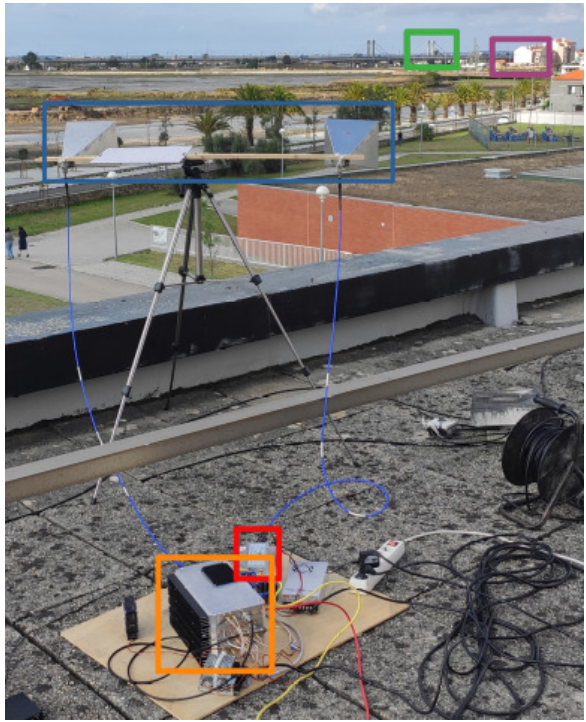

Fig. 4. Photo of the real setup. Each component in the diagram of Fig. 3 can be identified by the matching colored boxes.

### 3.2 Waveform Design and Signal Processing

The pulse waveform used in this test was an AM Chirp. Chirp waveforms are very popular in radar systems due to highly compressed autocorrelation and Doppler resilience [11]. ATLAS has amplitude modulation so it is not possible to create conventional chirps with it. The AM Chirp consists of a modified chirp signal at a much lower frequency used for modulating the carrier in amplitude [4]. The AM Chirp signal, y(t) is given by:

$$y(t) = A(t)\cos(2\pi f_c t) \quad (1)$$

$$A(t) = \alpha + \beta\cos[2\pi(ct + f_0)t], \; c = \frac{f_1 - f_0}{T} \quad (2)$$

where $f_c$ is the carrier frequency, $\alpha$ and $\beta$ define the modulation parameters, $f_0$ and $f_1$ are the starting and final frequencies, respectively, and T is the pulse duration. The AM chirp used in the tests had the following parameters: $\alpha = 0.5$, $\beta = 0.5$, $f_0 = 1$ MHz, $f_1 = 3$ MHz and T = 160 us. After pulse compression, the developed AM chirp presents a range resolution of 150 m at 5.56 GHz (ATLAS operating frequency).

In order to detect the presence of echoes in an extremely low SNR context, some signal processing needs to be applied. Fig. 5 depicts the signal processing chain implemented in order to increase the detectability of the targets. The system was configured to generate a file for each emitted pulse that contains all the echoes received from that pulse in a 5 km range.

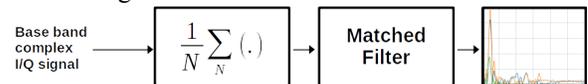

Fig. 5. Signal processing chain.

In each measurement, 500 pulses were emitted which resulted in 500 files with complex I/Q data. The signal processing chain coherently integrates the 500 files in order to reduce white gaussian noise. After integration, the signal goes through a matched filter. After proper normalization, delay compensation and time to distance conversion, the output consists in a data series of peaks dependent on the distance. This can be passed through a CFAR algorithm for automatic detection [12] or can be visually inspected to detect the presence of targets.

### 3.3 System Performance

The complete system characteristics, from the radar hardware capabilities to the type of antennas, the waveform design and the signal processing techniques used will have an impact on the system performance. The radar equation is a useful tool to predict how the system will perform given all the factors aforementioned [13]. The following form of the radar equation provides the dependence of SNR with the system specifications:

$$SNR = \frac{P_{av} G^2 \lambda^2 \sigma\, ne(n) F^4}{4\pi^3 \tau f_p R^4 N_F kTBL_s} \quad (3)$$

Table 2 has a brief description of all the variables in equation 3 as well as the values for the end to end test.

Table 2. Radar equation parameters

| Variable | Description | Value |
|---|---|---|
| $P_{av}$ | Average transmitted power (mW) | 24 |
| G | Antenna Gain (dBi) | 17 |
| λ | Operating Wavelength (cm) | 5.4 |
| σ | Target RCS (m²) | 1 |
| n | Number of integrated pulses | 500 |
| e (n) | Integration Efficiency | 1 |
| F | Propagation Effects factor (dB) | 0 |
| τ | Pulse Width (µs) | 160 |
| $f_p$ | Pulse Repetition Frequency (Hz) | 41.7 |
| R | Distance to the target (km) | [0...5] |
| $N_F$ | Noise factor of the receiver (dB) | 0.7 |
| k | Boltzmann constant (J/K) | 1.38e-23 |
| T | Receiver Temperature (K) | 288 |
| B | Receiver Bandwidth (MHz) | 2 |
| $L_s$ | System Losses factor (dB) | 11 |

Even though the system uses two antennas, they are identical and have the same gain. The test did not include a standard target with a well defined RCS so there was no control over the RCS of the targets, the system performance assessment was computed considering a spherical 1 m² target. As shown in Table 1, the radar receiver has 80 MHz of bandwidth, but using a matched filter reduces the bandwidth of the signal to the bandwidth of the waveform, which is 2 MHz. Given the values in Table 2, Fig. 6 depicts the SNR dependence on the distance.

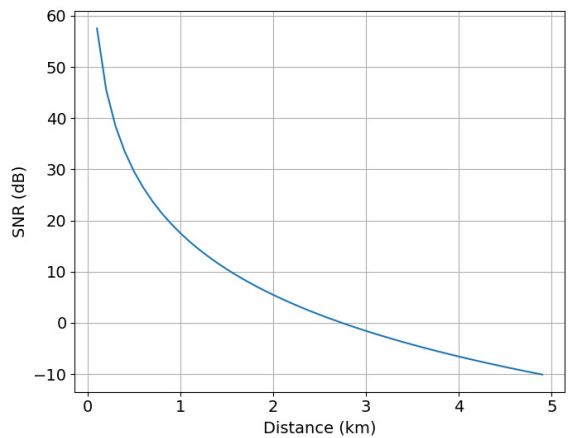

Fig. 6. Predicted SNR dependence on distance for the end-to-end test setup.

## 4. Results and Discussion

The experiment consisted of detecting a well defined target with high reflectivity that was present in the landscape around IT. To do that we mounted the setup as depicted in Fig. 4 and pointed the antennas to the triangular structures in the highway that are highlighted in the green box of Fig. 4. We then performed two experiments: emitting two bursts of 500 pulses to the target and emitting the same burst but pointing to the sky. The first two trials were identical but spaced in time by around 10 minutes with the objective of checking if the measurements were consistent, pointing to the sky was also to have a comparison between the signal received by pointing to reflecting targets or to the sky, which we assume had no relevant reflectors in a range of 5 km.

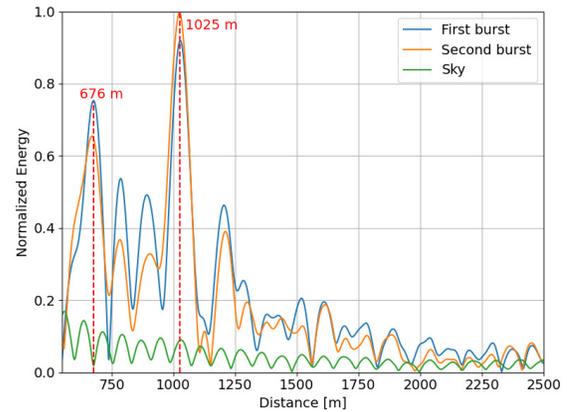

Fig. 7. Range profiles obtained for 3 experiments.

For each experiment, the data was collected and passed through the processing chain (Fig. 5) which generates a range profile. Fig. 7 shows the range profiles obtained for each experiment.

The first information we can easily gather from these results is the absence of peaks on the sky experiment in comparison to the other two. We can safely assume that the peaks obtained when pointing to the targets are indeed reflections from the emitting pulses since we did not obtain any signal when pointing to the sky, given that the likelihood of randomly pointing at the sky and finding an object is low.

Comparing the first and second burst experiments, the two profiles are similar, with two main peaks in the same location, which confirms that the peaks are consistent from experiment to experiment and are in fact highly reflecting targets. The targets are most likely stationary since they maintained the same distance to the radar in the two experiments.

The range profiles exhibit two main peaks, one at 676 m and another at 1025 m. The distance between the setup and the target depicted in Fig. 4 (triangular structures) is roughly 1020 m and was measured using Google Maps distance measurement tool. Keep in mind that the peaks have a resolution of 150 m so all objects inside that range contribute to the same peak. Said that, the peak at 1020 m can be given not only by the triangular structure itself but to other nearby objects not possible to resolve.

The peak obtained at 676 m is also very predominant in the range profile, it was not intentional since we did not intend to point to any structure at that distance. At this distance the radiation has spread around 140 m, given the 12º of beamwidth of the antennas. Considering the beamwidth spread and the range resolution of 150 m, it is possible that the energy on this peak corresponds to reflections from the buildings and structures on the dark purple box in Fig. 4.

**5.    Conclusions and Future Work**

ATLAS is one of the first steps in Portugal to the establishment of a tracking radar system that provides orbital object measurements for SST activities. It is located at the ErPoB station, a facility in the center of Portugal Mainland away from major RF interference, that houses all the equipment and is accessible remotely through a secure connection to IT- Aveiro, where the main research and development activities are performed. The radar system features cutting edge hardware technology using a highly modular architecture with full digital processing. It establishes a modern and versatile platform for fast and easy development, research and innovation.

The whole system (except antenna and power amplifiers) was tested in a setup with a major reflector at a well defined distance. The obtained range profiles show that the target can be detected easily.

Now that the system was tested from emission to reception and consequent signal processing, the next steps consist of doing similar measurements to more distant objects by using better power amplifiers and finally deploying it at the ErPoB to detect objects in orbit.

**Acknowledgements**

This paper and team work has been supported by the European Commission H2020 Programme under the grant agreement 2-3SST2018-20; The team acknowledges further support from ENGAGE-SKA Research Infrastructure, ref. POCI-01-0145-FEDER-022217, funded by COMPETE 2020 and FCT, Portugal; exploratory project of reference IF/00498/2015 IT team members acknowledge support from Projecto Lab. Associado UID/EEA/50008/2019.

The team acknowledges LC Technologies for technical developments in close collaboration with IT.